# INCREASE OF SOFTWARE SAFETY

**Arkadiy Khandjian,** hark2004@mail.ru
**Moscow, 01-07-2008**

New model of software safety is offered. Distribution of mistakes in program on stages of life cycle is researched. Study of ways of increase of reliability of software at help simulation program is leaded.

## INTRODUCTION

Using of software on many objects requires study of question of increase of reliability of such software. Each mistake can bring in software on such objects to serious consequences and even distresses. In addition conciseness of periods required for development, limitation does not often allow in human and financial resource to reach required parameters of software safety. So it is necessary to develop recommendations on creation of reliable software, forecasting of features of software in condition scarce resource and achievement of required parameters of software safety.

## 1. RELIABILITY MODEL ON BASE OF MARKOV QUEUE SYSTEMS

### 1.1. Model of errors appearance and purification

We will consider appearance and purification in program how Markov death process and duplications with continuous time and we will find its features. Intensity of entering of mistakes in program in result completions, improvements and proofreading is equal $\lambda(t)$. Each mistake brought in program is found and corrected through random time $T$, distributed on significant law with parameter $\mu$ (time is distributed on significant law, as it is expected that this is simple stream of events with forgetting). We will consider stochastic process $X(t)$ - number of mistakes in program in instant $t$. We will find one-dimensional law of distribution of stochastic process $X(t)$. In the work [ 6 ] it is shown that general solution at starting condition $m_x(0)$ will be:

$$m_x(t) = \exp\left(-\int_0^t \mu(\tau)d\tau\right) \cdot \left[\int_0^t \left\{\lambda(x) \cdot \exp\left(\int_0^x \mu(\tau)d\tau\right)\right\}dx + m_x(0)\right] \qquad (1)$$

In compliance with problem specification it is necessary to decide this equation at starting condition $m_x(0) = X(0) = N$ - quantity of mistakes in program in initial time.

At constant rates $\lambda = const$ and $\mu = const$ equation (1) type will take:

$$m_x(t) = \frac{\lambda}{\mu} \cdot \left(1 - e^{-\mu t}\right) + N \cdot e^{-\mu t} \qquad (2)$$

**1.2. Distribution of mistakes on stages of life cycle**

We will consider simplistic example of software work. Intensity of streams of destruction $\mu$ and duplications $\lambda$ mistakes on different stages of life cycle of software they are shown in the picture:

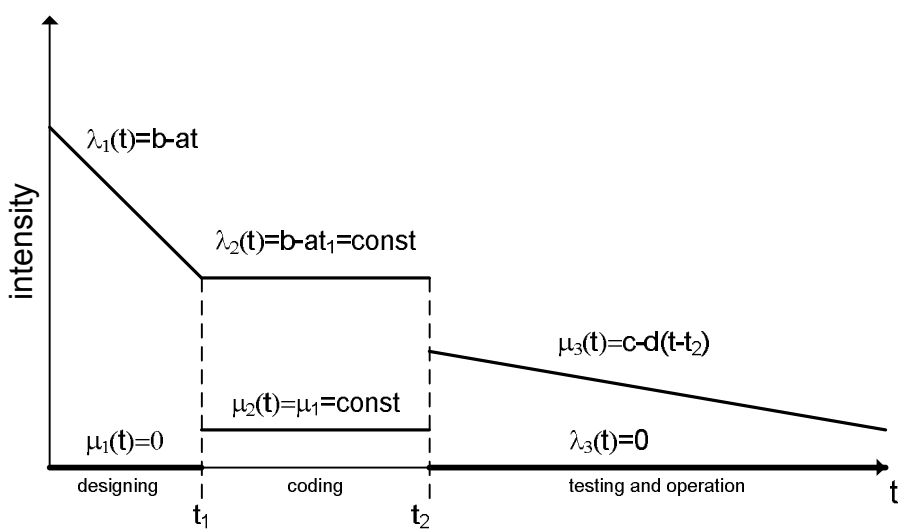

**Figure 1 - Intensity of streams of destruction and error-spread effect**

On first stage ($0 \leq t \leq t_1$) - stage of designing.   This is process of "pure" error-spread effects (verification of this stage can give stream of destruction of mistakes in end of this stage, but I consider that its contribution not essential).

On second stage (stage of coding $t_1 \leq t \leq t_2$) error flow is stabilized and depend mainly from those error sources, that were pawned on previous stage. At this point process of debugging also occurs, so each mistake can be corrected.

On third stage (operation phase and testing $t > t_2$) pure process of destruction of mistakes exists, moreover to rectify errors then all becomes developers of the development more difficult because of forgetting. In addition we consider that new mistakes are not given rise. We will determine expectation $m_x(t)$ numbers of program errors, if on moment of beginning of software designing $t=0$ mistakes were not in him ($m_x(t) = D_x(t) =0$).

Solution of equation (1, 2) for each stage give graph to dependence $m_x(t)$, shown in the picture:

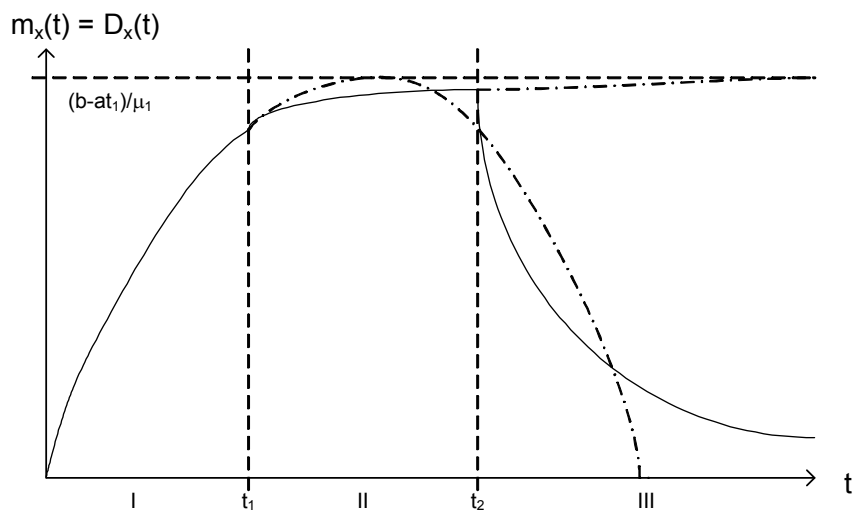

**Figure 2 – Quantity of mistakes in software during the life-cycle**

From graph one can see that it is useless to increase time of development of (coding) $t_2$ indefinitely, as quantity of mistakes will not anyway decrease. And only on third stage it is possible to reach essential decrease of quantity of mistakes, when new mistakes are not practically brought in program, and old are quickly corrected.

## 1.3. Development of model of reliability of client programs in software of client-server type

Far with help method of dynamic of average (see for example [ 1 ]) we will build Markov model of behaviors of program of consisting of many (approximately same) modules or (what is now used most often) we will build model of program system of type client-server. Characteristic feature of such a system is start server of parallel same streams, each of which services requests of one program-client or work of server with many same client programs. In that case streams or program- clients are completely identical and each of them can fail independently from other. *Feature of this system in difference from systems of considered in queueing theories (for example, service repair crew of car, or same hardware complexes) it is concluded in that, that at failure (finding mistakes) in one module (stream or client) and removal this mistake, this mistake is automatically avoided also in all another units (streams), as these streams are duplicated with way of start on performance of the same program* *code.* We will consider this feature at using of method of dynamic of average. In addition on change of module with mistake on corrected module we neglect with time.

So, let complex is (type client -- server) program system *S*, consisting from big number of homogeneous modules (streams or clients) *N*, each of which can with at random go from condition able. Let (for simplicity) all event flows (in case program - this is streams of external datum or requests from client programs to server), translating system *S* and each of its module) from condition able is Poisson (can be even with intensities, depending from time). Process, running in system, will then be Markov.

We will spread model on case most often met on practice, when each module-client is in one of two conditions: worker or non-working.

Let system *S* consists from large number *N* homogeneous elements (modules or streams of one unit), each of which can be in one of two conditions: $\xi_1$ - is efficient (work); $\xi_2$ - not worker (mistake is found and it is corrected).

On each module stream of mistakes with intensity acts $\lambda$, that depends from quantity of corrected more early in module of mistakes. Each defective component is corrected in average at the rate $\mu$ in unit of time. In initial moment (*t = 0*) all the elements (modules) they are serviceable. All event flows is Poisson (can be with variable intensity). We will write equations of dynamic of average for average numbers of conditions. Graph of conditions of one unit has type, presented in the picture:

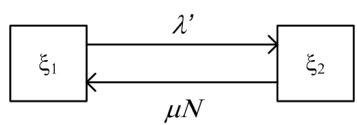

**Figure 3 - Graph of condition of module**

Here $\lambda$` -- intensity of stream of mistakes subject to previous corrections.

Find $\lambda$` from number of previous corrections of this module. We will state suggestion that $\lambda$` fall with quantity of corrected mistakes to some constant value of tolerance to corrections (for example, when quantity of corrected mistakes becomes equal to quantity of brought mistakes, or quantity of mistakes in module becomes so small, that they begin to work with constant rate) also to strive on exponential law to some minimum the more quickly, mistakes are the more quickly corrected $\mu$.

On base of graph (see Figure 3 - Graph of condition of module) differential conditions of dynamic means will enter the name in the form:

$$\begin{cases} \dfrac{dm_1(t)}{dt} = -\lambda' \cdot m_1(t) + \mu \cdot N \cdot m_2(t) \\ \dfrac{dm_2(t)}{dt} = -\mu \cdot N \cdot m_2(t) + \lambda' \cdot m_1(t) \end{cases}$$

where $m_1(t)$, $m_2(t)$ – Average number of conditions $\xi_1$ and $\xi_2$.

Of these two equations it is possible to choose one - for example, second, and to reject first. In second equation we will substitute expression for $m_1(t)$ from condition: $m_1(t) + m_2(t) = N$.

In addition quantity of modernizations $m$ depend from intensity of correction of module $\mu$ and quantities of programmers (or groups of programmers) $P$ working above correction of modules. We will expect that: $m(\mu) = \mu \cdot P \cdot t$ and $\lambda' = \lambda_0 \cdot e^{-\mu \cdot P \cdot t} + \lambda_1$.

It is necessary to decide this equation at starting condition $m_2(t=0) = 0$ of numerical methods. This equation was decided with help package of mathematical programs MatLab 6.5 methods Runge - Kutt (function *ode45*) and following result is got for problem specifications: $\mu = 0,2$ times in day one mistake is corrected by one programmer; $\lambda_0 = 10$ times are found in day in program mistakes in initial time; $P = 3$ - quantity of programmers (or groups of programmers), correcting mistakes with intensity $\mu$ each; $N = 10$ - quantity of modules (streams or clients) in programs of client-server type; $\lambda_1 = 0,1$. We get decision, shown in the picture:

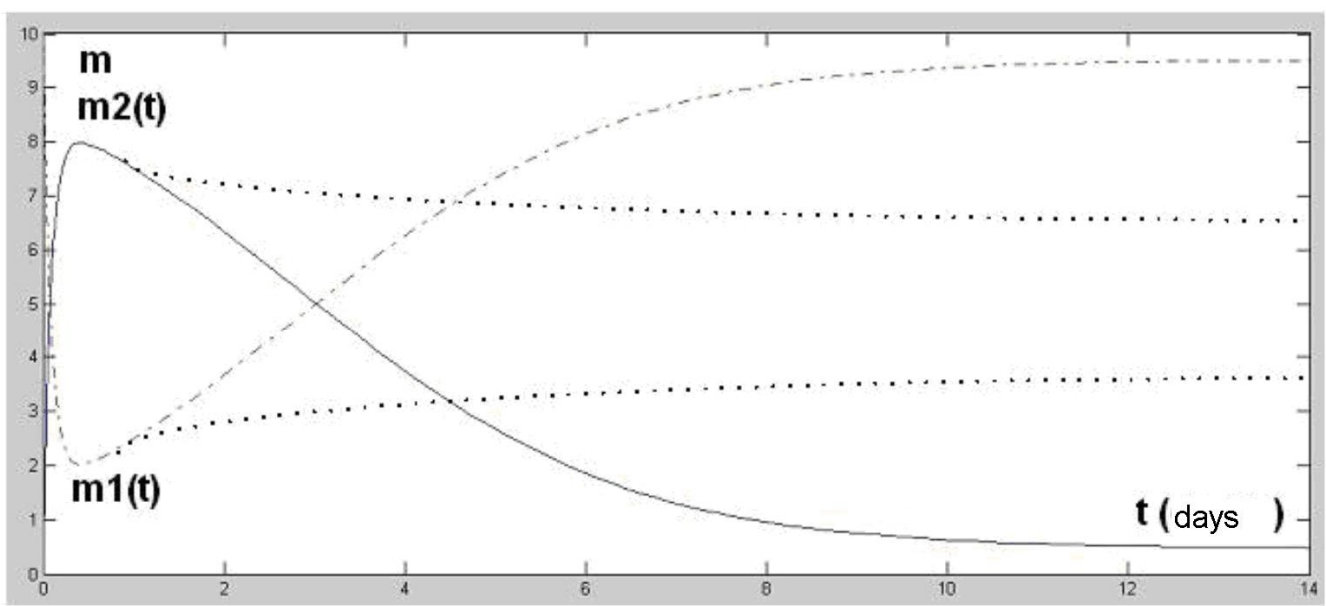

**Figure 4 - Decision for exponential dependence**

Explanations to figure: continuous line - $m_2(t)$ - number of idle modules; dash line (point of dash) - $m_1(t)$ - number of working modules; intermittent lines of (light dot) - representation of behaviour of crooked at absence of dependence $\mu' = N \cdot \mu$ -- features inherent of software. One can see that in that case time behaviors occur much more slowly.

From figure one can see that quantity of working modules will exceed quantity of idle modules on *3* days. Since this moment it is possible to consider that program work steadily. If originally $\lambda_0 = 100$, then at the same other starting conditions program began steadily to work only on *7* days.

In technology, where $\mu$ are not proportional $N$, at given starting conditions all technology would be faulty already on first day and would be never brought operational. So, without looking on that, that software in beginning of operation or (testing) contains in abundance (much more than in technology) mistakes with big intensity of their display, mistakes are quickly corrected in programs and very quickly software become firm. This occurs because of double-quick correction of mistakes as contrasted to technology.

**1.4. Development of general reliability model of software of client-server type how Markov model of mixed mode**

### *1.4.1. Formulation and finding of basic formulas*

We will consider equations of mixed mode now. Heretofore we described processes, running in software, or with help equations for state probabilities, or with help equations of dynamic of average, where unknown functions are averages of number of conditions. Equations of first type are used then, when software is comparative simple and its conditions are rather small. Equations of second type are specially reserved for description of processes, occurring in software, consisting of multiple modules. For such a systems we managed to find not state probability, and average number of conditions.

On practice situations of mixed mode are more often met. Also for such software we will write equations. This model is applicable for software, that consists of element- modules of different type: small (unique) (for example, in architecture client-server this - server) and multiple (in architecture client-server this - clients), moreover conditions of those and others are relative.

In this case for modules of first type it is possible to make differential equations, where unknown functions are state probabilities. For modules of second type - average number of conditions. We will call such equations call equations of mixed mode.

We will consider software $S$, consisting from plenty $N$ identical client programs and one server, that coordinates work of all client programs. How server, and separate clients can refuse (to hang). Intensity of stream of refusals of server depends from number $x$ of working program- clients (that is actually depend from intensity of input datum and their range): $\lambda^c = \varphi(x) = \lambda'$. Intensity of stream of faults of each module- client at working server is equal $\lambda$`.

Average time of removal of mistake in server, taking into account difficulty of server, more that average time of removal of mistake in client: $\bar{t}^c_{ucnp} = S / \mu_0$, where $\mu_0$ -- speed of removal of mistakes in client (speed of error correction programmer), $S$ - factor of difficulty of server.

We will describe process, running in program provision with help equations of mixed mode, where they will be unknown functions: probabilities of conditions of server; average number of conditions of clients. We will describe our system at help column, shown in the picture:

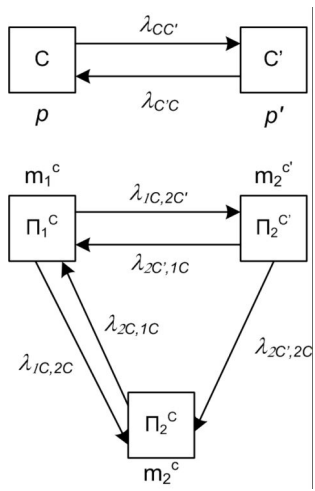

**Figure 5 - Graph of combination system**

This graph breaks up on two subgraph. First (top) - this is subgraph of conditions of server, that can be in one of two conditions:

$C(t)$ - work; $C'(t)$ - do not work (mistake is found and it is corrected).

What touches program- client, then for her we take into account opportunity to be in one of three conditions: $\Pi_1^C(t)$ - client works at working server; $\Pi_2^C(t)$ - client does not work at working server; $\Pi_2^{C'}(t)$ - client does not work at not working server.

Condition of server is characterized in instant $t$ one of events $C(t)$ and $C'(t)$. We will mark probabilities of these events through $p(t)$ and $p'(t) = 1 - p(t)$, and number of conditions $\Pi_1^C(t)$, $\Pi_2^C(t)$ and $\Pi_2^{C'}(t)$ accordingly: $X_1^C(t)$, $X_2^C(t)$ и $X_2^{C'}(t)$.

It is obvious, for any instant $t$: $m_1^C(t) + m_2^C(t) + m_2^{C'}(t) = N$         (3)

where is $N$ - number of clients, working with server.

We will determine intensity of streams of events for graph (see Figure 5). Before all, under the statement of problem:: $\lambda_{CC'} = \lambda^C = \varphi(X_1^C) \approx \varphi(m_1^C) = \lambda'$, $\lambda_{C'C} = \dfrac{1}{\bar{t}_{ucnp}^C} = \dfrac{\mu_0}{S}$.

Far, program-client goes from condition $\Pi_1^C(t)$ in condition $\Pi_2^{C'}(t)$ not themselves on themselves, and only together and it is simultaneous with server (when that hangs). So:

$$\lambda_{1C,2C'} = \lambda_{CC'} = \varphi(X_1^C) \approx \varphi(m_1^C) = \lambda'. \text{ It is similar: } \lambda_{2C',1C} = \lambda_{C'C} = \frac{\mu_0}{S}.$$

For other transitions it is not difficult to establish corresponding to intensity, if to consider that fact that vote (low) subgraph is distinguished from considered more early (see Figure 3) only presence of still one condition $\Pi_2^{C'}$, when client program stands on time of correction of mistake in program-server. With account this we have: $\lambda_{1C,2C} = \lambda' = \lambda_0 \cdot e^{-\mu P_d} + \lambda_1$; $\lambda_{2C,1C} = \mu_0 \cdot N$; $\lambda_{2C',2C} = \lambda_{2C',1C}$.

We will write for graph (see Figure 5) differential equations of mixed mode, approximately describing our system (argument $t$ are lowered for brevity of record):

$$\begin{cases} \dfrac{dp}{dt} = -\lambda_{CC'} \cdot p + \lambda_{C'C} \cdot p' \\[2mm] \dfrac{dp'}{dt} = -\lambda_{C'C} \cdot p' + \lambda_{CC'} \cdot p \\[2mm] \dfrac{dm_1^C}{dt} = -\left(\lambda_{1C,2C'} + \lambda_{1C,2C}\right) \cdot m_1^C + \lambda_{2C',1C} \cdot m_2^{C'} + \lambda_{2C,1C} \cdot m_2^C \\[2mm] \dfrac{dm_2^C}{dt} = -\lambda_{2C,1C} \cdot m_2^C + \lambda_{1C,2C} \cdot m_1^C + \lambda_{2C',2C} \cdot m_2^{C'} \\[2mm] \dfrac{dm_2^{C'}}{dt} = -\left(\lambda_{2C',1C} + \lambda_{2C',2C}\right) \cdot m_2^{C'} + \lambda_{1C,2C'} \cdot m_1^C \end{cases} \qquad (4)$$

We will note that, having placed in (4) all first members $0$, it is possible to puzzle out for stationary condition, and he exists, as system ergodic.

We will notice that from this set of equations it is possible to exclude two equations: one of first two, using equation $p + p' = 1$, and one - from following three, using parity of normalization (3). These equations are decided on the assumption, that at the beginning server and all program- clients work: $t = 0; p = 1; p' = 0; m_1^C = N; m_2^C = m_2^{C'} = 0$.

### 1.4.2. Example of model use

We will allow set of equations (4) for following conditions: $S = 3$ - factor of difficulty of server; $N = 10$ - number of program- clients; $\lambda_0 = 10$ mistakes / day; $P = 3$ - quantity of programmers;

$\mu_0 = 0,5$ mistakes / day. Decision of foregoing model was leaded with help package *MatLab6.5* (function *ode15s*) method of Runge - Kut.

Following results are got:

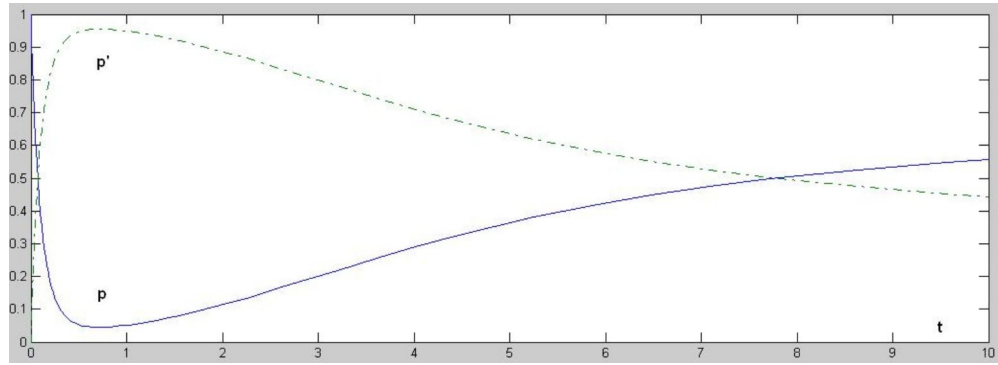

**Figure 6 – p** and **p'**

From figure one can see that server will begin steadily to work on *8* days.

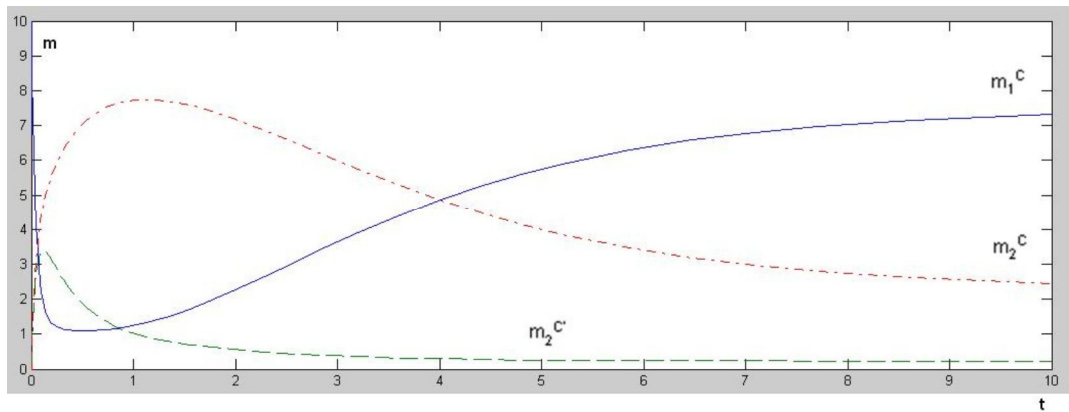

**Figure 7 – m$_1^C$, m$_2^C$, m$_2^{C'}$**

From figure one can see that clients will begin steadily to work on *4* days.

If to increase quantity of program- clients with *10* to *100*, then results will not practically change or output will occur on operational stability even more quickly. If to reduce number of programmers from 3-x people to one, then server will begin steadily to work on *14* days, and clients - on *10* days.

# 2. STUDY OF WAYS OF INCREASE OF RELIABILITY OF SOFTWARE ON BASE OF OFFERED SOFTWARE MODEL

## 2.1. Formulation

Master problem of presence of reliability of software at help many reliability models is need to know initial quantity of mistakes in program provision. Regrettably, offered reliability model does not allow to find this value. She does not in general use her. That not less, results got at use of this model well are in agreement with with practice. So it is possible to try to use these results for finding $N_0$ - initial quantity of program errors from method of back calculation. This will let use also other reliability models. Such simulation program will also let find such features of software safety, how time to following refusal, its probability and time of achievement of necessary reliability at given starting conditions. Also such simulation program will let research to way of increase of software safety, varying one of available in order of developers resource, such as quantity of programmers, speed of purification, speed of entering of mistakes, time of testing and intensity of testing.

Following candidate solution is offered to this problems: with allowance for more early got results about character of behaviour of reliability of program provision from time (Poisson distribution; mixed model of type client-server, when server is faulty, then clients stand: simultaneous correction of mistake in all clients at correction of mistake in some of one client) to write program of modeling of behaviour of software safety (modelling process of finding of mistakes in program provision and purifications) and to select $N_0$ with its help with hence, so that final features of software safety coincided with results, got at help more early offered reliability model.

## 2.2. Description of operation of simulation program

Program provision of type is client-server. Server services requests from $N$ program- clients (far simply clients). In program provision is even on area of definition of input datum (ADI) (A, B) they are arranged $Er$ mistakes. Server in a complicated more way than program- clients from point of sight of development of software in $S$ time. $S$ - factor of difficulty of server with respect to clients. Each $k$-client ($k = 1, 2, ..., N$) generates Poisson stream of data to server with intensity $\lambda_{обр}$. Data from the client are distributed on ADI on normal law with features $m_k$ and $\sigma_k$, where $m_k$ are distributed between clients evenly on all area of input datum, $3\sigma_k$ - it is distributed evenly on small of sites of cut $m_k$ on axes of data area. This is necessary for imitation of unevenness of use ADI at small quantity of clients.

On request of client server answers with data, that are distributed evenly on all definition domain (A, B).

In the picture (see Figure 8) distribution of requests of one is represented client on area of all possible requests to server, and also even distribution of mistakes on ADI is shown. At hit of request of client or answer of server in area ADI, containing mistake, it is considered that mistake is found and appropriate module is decommission for its correction:

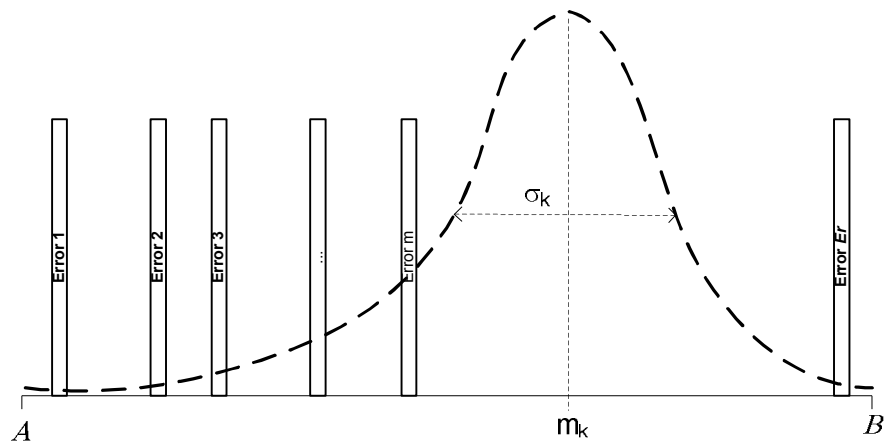

**Figure 8 - Distribution of requests k-client on data areas**

For modeling of streams of destruction and duplication of mistakes in program provision is used Monte-Carlo methods.

Input datum for drawing they are: $P$ - quantity of programmers, servicing system; $K$ - quantity of program- clients; $\alpha$; - width of one request of client how share from ADI (from $0$ to $1$, where $1$ - this is all ADI); $\Delta t$ - step of iteration (day); $s$ - factor of difficulty of server as contrasted to program- client; $\lambda_{o\delta p}$ - intensity of stream of addresses of one client to server (1 / day); $\lambda_{ucnp}$ - intensity of stream of error correction one programmer (1 / day); $\lambda_{\text{внес}}$ - intensity of entering of mistake at correction one programmer (1 / day) or $p_{\text{внес}}$ - probability to bring mistake at correction by one programmer; $M$ - quantity of iterations (quantity of attempts of addresses of program- clients to server one drawing); $R$ - quantity of drawings for averaging; *Er* - initial quantity of mistakes. Program text and performed program are placed on site of www.arkpc.narod.ru.

## 2.3. Practical simulation results

### 2.3.1. Influence of quantity of clients on software safety

We will study influence of quantity of program- clients on behaviour of software. Drawing was conducted at following initial conditions of (*10* clients):

*Number of program-clients: 10, Number of programmers: 3, Share from general data area (ADI) in one request of client: 1E-5, Initial number of mistakes: 250, Factor of difficulty of server: 2, Intensity of stream of addresses of client to server: 500 (1 / day), Intensity of stream of error correction: 1 (1 / day), Intensity of entering of mistake at correction: 0,1 (1 / day), Step Of Iteration: 0,002, Number of iterations: 50000, General time of drawing: 100 (day); Number of drawings: 50*

Following results are Got (average values for all 50 draws, see Figure 9 -- Drawing № 1):

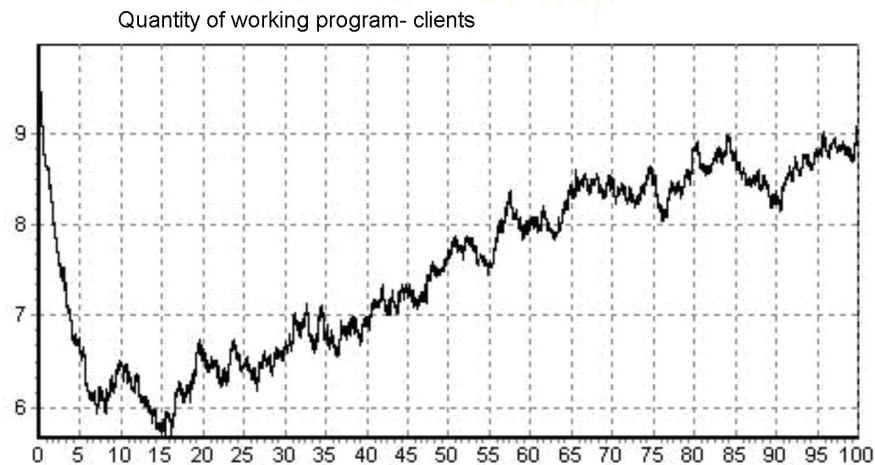

**Figure 9 – Drawing № 1**

From figure one can see that software will begin steadily to work (i. e. quantity of working clients) will moved into line with quantity of idle clients on *15* days, what well is in agreement with computational model, see Figure 6 and 7.

We will now increase quantity of clients with *10* to *100*:

Following results are Got (average values for all *50* draws, see Figure 10 - Drawing № 2):

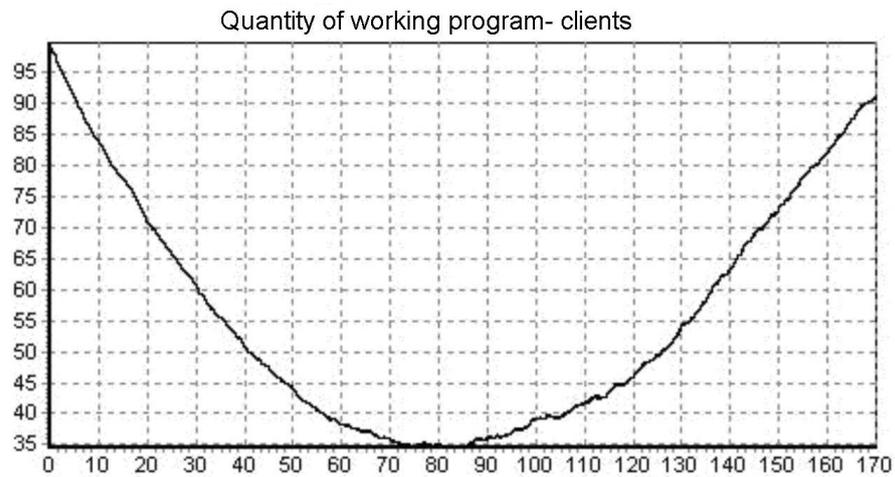

**Figure 10 - Drawing № 2**

One can see that on 170 days all mistakes are almost corrected. This occurs because of that, that clients more and their requests cover big data area and, hence, great quantity of mistakes is found and great quantity of mistakes is corrected. At ten clients in program provision will still remain on 170 days about 50 mistakes.

### 2.3.2. Influence of quantity of programmers on software safety

We will now show that at light load on server (small quantity client programs) increase of quantity of programmers, correcting a mistake, gives small effect. Quantity of uncorrected errors to end of testing remains the previous. Only time of waiting of program of correction in turns decreases. For example, if to increase quantity of programmers with 3 to 12, that we will get:

Initial conditions of drawing:

*Number of program-clients: 10, Number of programmers: 12, Share from general data area (ADI) in one request of client: 1E-5, Initial number of mistakes: 250, Factor of difficulty of server: 2, Intensity of stream of addresses of client to server: 500 (1 / day), Intensity of stream of error correction: 1 (1 / day), Intensity of entering of mistake at correction: 0,1 (1 / day), Step Of Iteration: 0,002, Number of iterations: 50000, General time of drawing: 100 (day); Number of drawings: 50*

Results of drawing:

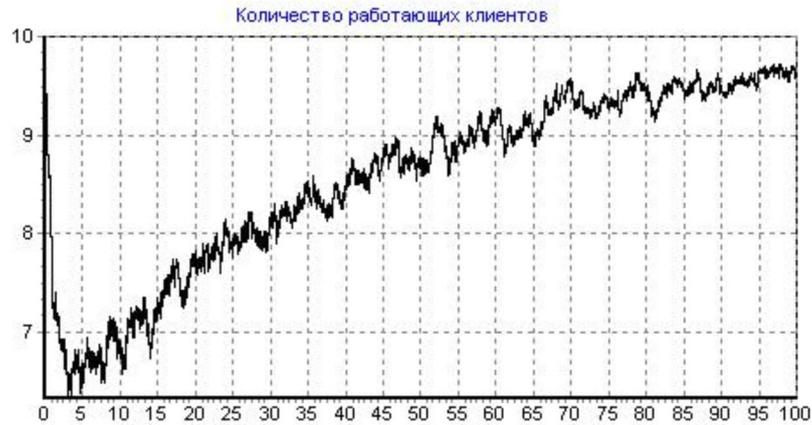

**Figure 11 - Drawing № 3**

One can see that program will begin it is firm to work how and earlier only on 10-15 days, then not big effect lets eat increase of quantity of programmers and sooner than all, part of programmers will stand. Much more effectively to increase load in this situation at testing. How for example, this was already shown more highly, increasing quantity of clients. Increase of quantity of programmers can render even adverse effect on software safety, if at removal of mistakes in program they intensively bring provision in him new mistakes. We will show this on example.

Let at 12 programmers each of them brings mistake with intensity 0,6 instead of 0,1 mistakes in day.

Results of drawing:

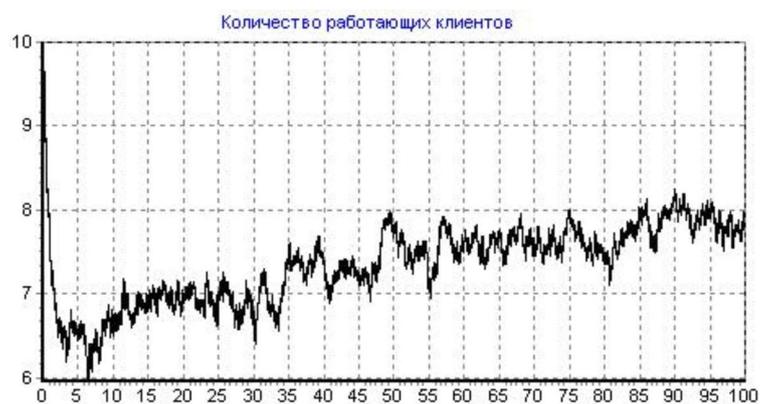

**Figure 12 - Drawing № 4**

From figure one can see that for 100 days of system operation quantity of mistakes did not practically decrease.

### 2.3.3. Influence of intensity of addresses of clients to server

Increasing intensity of address of each client to server does not give such effect, because each client usually works in the narrow part of data area and dislodges mistakes from this part, and significant remains data area not checked, and mean in errors. Example of drawing at increases of intensity of addresses on order with 500 to 2500 in day.

Initial conditions of draw:

*Number of program-clients: 10, Number of programmers: 3, Share from general data area (ADI) in one request of client: 1E-5, Initial number of mistakes: 250, Factor of difficulty of server: 2, Intensity of stream of addresses of client to server: 2500 (1 / day), Intensity of stream of error correction: 1 (1 / day), Intensity of entering of mistake at correction: 0,1 (1 / day), Step Of Iteration: 0,0004, Number of iterations: 250000, General time of drawing: 100 (day); Number of drawings: 10*

Results of drawing:

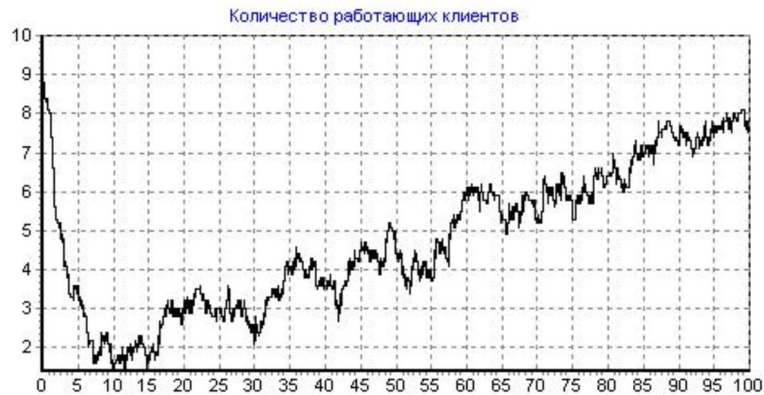

**Figure 13 - Drawing № 5**

### 2.3.4. Definition of initial quantity of mistakes in program provision

Given model allows with a combination with offered Markov model of software safety to value quantity of program errors with in the following way - to get accounting result, and then to select initial quantity of mistakes in program provision such, so that results of draw coincided with result of calculation.

For decision of this it is necessary to obtain problem with help simulation program that, so that initial intensity of error flow $\lambda_0$ from model of reliability of program provision of type client-server (see paragraph 1.4) coincided with initial intensity of stream of mistakes in simulation program. It is directly impossible to do this, as in program there is modeling of such parameter not. For that in simulation program it is necessary to place $\alpha = 0.5$, then there is each address of client to server and answer of server to client should with probability 1 to generate mistake. It is then necessary to obtain that, so that quantity of addresses for day of clients to server (i. e. K*$\lambda_{обр}$) was equal $\lambda_0$. It is necessary to place other initial parameters of simulation program with equal to similar parameters of reliability model.

We will find initial quantity of mistakes as an example considered in paragraph 1.4.2. In order that initial intensity of stream of mistakes in simulation program was equal $\lambda_0$=10 from example of paragraph 1.4.2, we will place $\alpha$ = 0.5, and $\lambda_{обр}$ at 3-th programmers we will place equal 3,3.

Initial conditions of drawing:

*Number of program-clients: 10, Number of programmers: 3, Share from general data area (ADI) in one request of client: **0,5**, Initial number of mistakes: 9, Factor of difficulty of server: 3, Intensity of stream of addresses of client to server: **3,3** (1 / day), Intensity of stream of error correction: 0,5 (1 / day), Intensity of entering of mistake at correction: 0 (1 / day), Step Of Iteration: 0,0001, Number of iterations: 100000, General time of drawing: 10 (day); Number of drawings: 50*

Results of drawing:

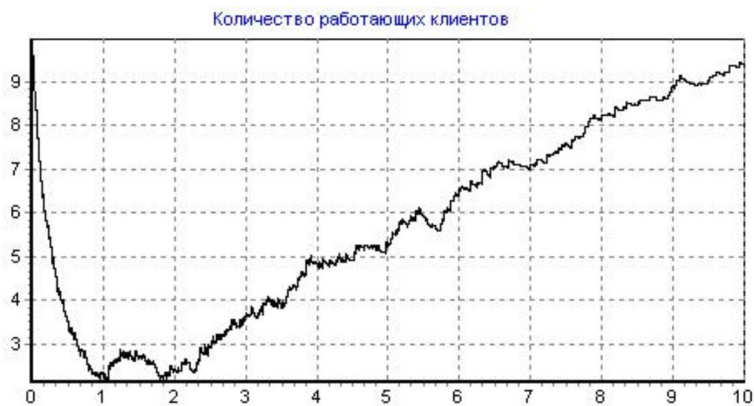

**Figure 14 - Drawing № 6**

How one can see from figure at initial quantity of program errors equal 9 got result similar got in model, then clients will begin to eat steadily to work on 4 days. Number 9 of *Er* were got with method of selection of different initial values of quantity mistakes in program on initial time.

Hence, combining model and drawing it is possible to calculate initial quantity of mistakes in program provision and other its features.

### 3. SUMMARY

1. New mathematical model of reliability of software on base of Markov queue systems, allowing to conduct calculation of features of software safety is built. Offered model more simpler than used more early than model. Master advantage of model is absence of use in her initial quantity of mistakes in program provision. Consideration of software how "black box" give consistent results, confirmed on practice without detailed description of all software performances.

2. Process properties are considered duplication and destruction of mistakes in program provision on different stages of life cycle. Master dependences of distribution of mistakes on stages of life cycle for many typical case are got.

3. For increase of software safety it is necessary to control two master making, influencing on software safety: before all - to raise intensity of testing or uses; and to raise quantity of programmers and / or efficiency of their work. In addition it is necessary to determine in condition limitation of resource and often taking into account uniqueness of development (software is often developed in single copy and for single unique object) how required software safety long should test software or conduct experienced operation of software in order to achieve. Estimation is given to time of achievement of required level of reliability of software at given quantity of programmers and their efficiency works. Probabilistic approach let to reliability give an answer on question of one of the most complex problems at testing: "When it is necessary to finish testing in order to satisfy requirements on reliability to software?"

4. For presence of optimum parity of features developments and software maintenance methodology of modeling of behaviour of reliability of software during is developed - program of modeling on base of Monte-Carlo method and based on offered to model of software safety are developed. On its base recommendations are developed for increase of software safety. Simulation program allows, setting different starting conditions, to observe behaviour of reliability of software during. This allows to value expenses and resource for building and support of highly reliable software. It is shown that dominant factor, allowing greatly to raise software safety, intensity of testing is.

5. Combination two approaches - Markov model of reliability of software and forecasting at help Monte-Carlo method - allow more exactly and more comprehensively to value features of software safety. Personallies, this allows to find initial quantity of mistakes in program provision.

**LITERAURE**